\documentclass[aps,reprint,floatfix,longbibliography]{revtex4-1}

\usepackage{amsmath}
\usepackage{graphicx}
\graphicspath{{figures/}}
\usepackage[separate-uncertainty=true]{siunitx}

\renewcommand{\vec}[1]{\mathbf{#1}}
\newcommand{\abs}[1]{\left\vert #1 \right\vert}

\begin{document}

\title{CATCH: Characterizing and Tracking Colloids Holographically using deep neural networks}

\author{Lauren E. Altman}
\author{David G. Grier}

\affiliation{Department of Physics and 
Center for Soft Matter Research,
New York University, New York, NY 10003}

\begin{abstract}
In-line holographic microscopy provides an unparalleled wealth of 
information about the properties of colloidal dispersions.
Analyzing one colloidal particle's hologram with the Lorenz-Mie theory 
of light scattering yields the particle's 
three-dimensional position with nanometer precision
while simultaneously reporting
its size and refractive index
with part-per-thousand resolution.
Analyzing a few thousand holograms in this way provides a comprehensive
picture of the particles that make up a dispersion,
even for complex multicomponent systems.
All of this valuable information comes at the cost of
three computationally
expensive steps: (1) identifying and localizing
features of interest within recorded holograms,
(2) estimating each particle's properties based
on characteristics of the associated features,
and finally (3) optimizing those estimates through
pixel-by-pixel fits to a generative model.
Here, we demonstrate an end-to-end
implementation that is based entirely on machine-learning
techniques.
Characterizing and Tracking Colloids Holographically (CATCH)
with deep convolutional neural networks
is fast enough for real-time applications
and otherwise outperforms conventional analytical
algorithms,
particularly for heterogeneous and crowded samples.
We demonstrate this system's capabilities with
experiments on free-flowing and holographically trapped
colloidal spheres.
\end{abstract}

\maketitle

\section{Introduction}
\label{sec:introduction}

Lorenz-Mie microscopy is a powerful
technology for analyzing the properties of colloidal particles
and measuring their three-dimensional motions
\cite{lee07a}.
Starting from in-line holographic microscopy images
\cite{sheng06,lee07},
Lorenz-Mie microscopy measures the
three-dimensional location, size and refractive index
of each micrometer-scale particle in the microscope's
field of view.
A typical measurement yields each particle's position
with nanometer precision over a hundred-micrometer range
\cite{cheong10a}, its size
with few-nanometer precision
and its refractive index to within a part per thousand
\cite{krishnatreya14}.
Results from sequences of holograms can be linked into
trajectories for flow visualization \cite{cheong09},
microrheology \cite{cheong08},
photonic force microscopy \cite{roichman08a},
and to monitor transformations in colloidal dispersions'
properties \cite{cheong09,shpaisman12,wang15,wang15a}.
The availability of \emph{in situ} data on particles' sizes,
compositions and concentrations 
is valuable for product development, process
control and quality assurance in such areas as
biopharmaceuticals \cite{wang16},
semiconductor processing \cite{cheong17},
and wastewater management \cite{philips17}.

Unlocking the full potential of Lorenz-Mie microscopy
requires an implementation that operates in real time and
robustly interprets the non-ideal holograms that
emerge from real-world samples.
Here, we demonstrate that this challenge can be met with
machine-learning techniques, specifically deep convolutional
neural networks (CNNs) that are trained with synthetic data
derived from physics-based models.

The analytical pipeline for Lorenz-Mie microscopy involves
(1) identifying and localizing features of interest in 
recorded holograms
and (2) estimating single-particle properties from the
measured intensity pattern in each feature
\cite{lee07a,soulez07,fung11,perry12,fung12}.
The CATCH network performs these analytical steps over
an exceptionally wide range of operating conditions,
yielding results more robustly and
\num{100} times faster than the best reference implementations based on
conventional algorithms \cite{cheong09,krishnatreya14a,dimiduk16,hannel18}.
The results are sufficiently accurate to solve
real-world materials-characterization problems 
and can bootstrap nonlinear least-squares fits
for the most demanding applications.

\section{Methods and Materials}
\label{sec:methods}

\subsection{Lorenz-Mie Microscopy}
\label{sec:lorenzmie}

\begin{figure}
    \centering
    \includegraphics[width=\columnwidth]{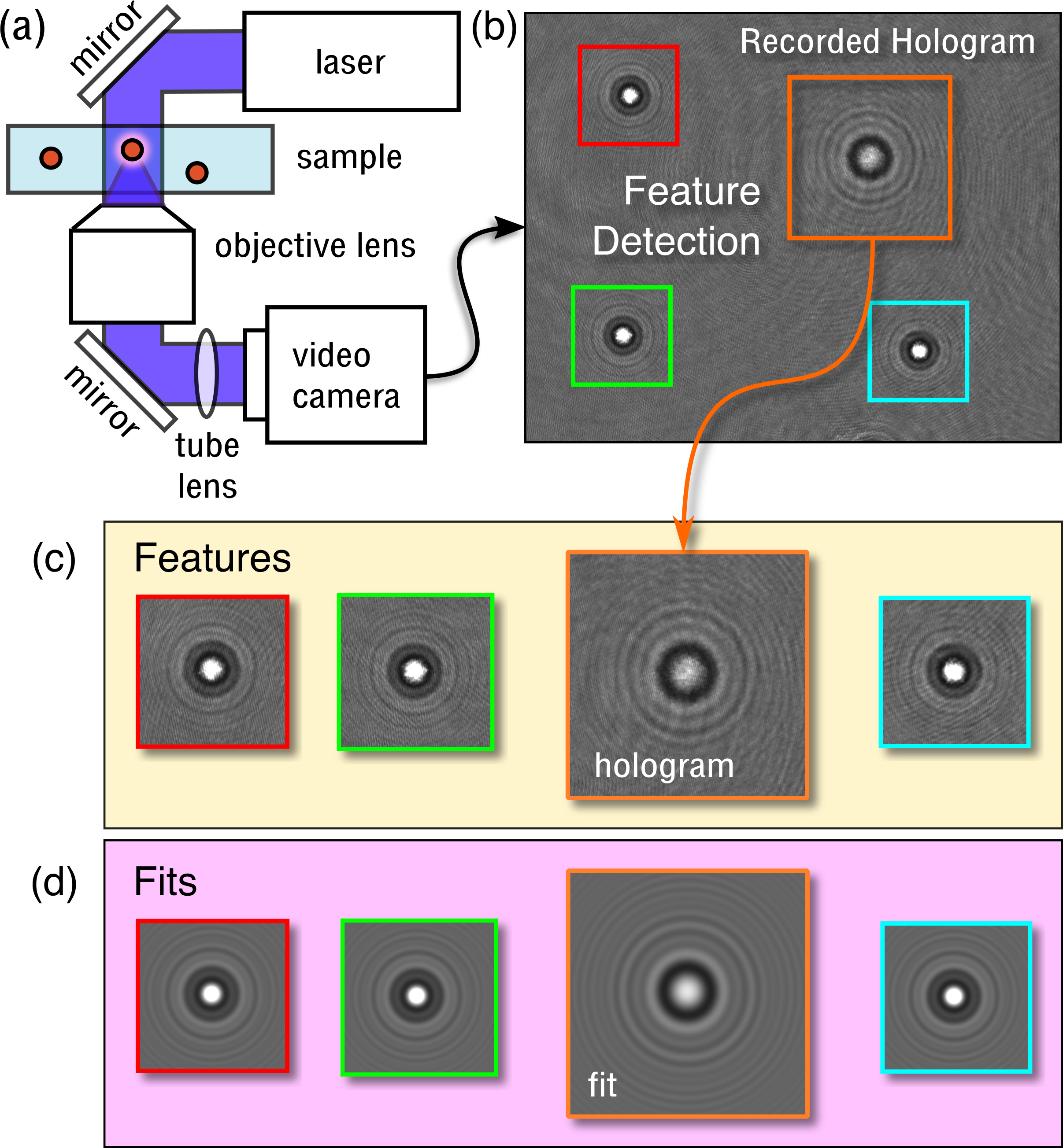}
    \caption{Schematic representation of Lorenz-Mie microscopy.
    (a) A fiber-coupled laser illuminates a colloidal sample.
    Light scattered by a particle interferes with the rest
    of the illumination in the focal plane of a microscope 
    that magnifies and relays the interference pattern to a video camera.
    (b) Each recorded hologram is analyzed to detect features of interest.
    (c) Each feature is localized within a region
    whose size is dictated by the local signal-to-noise ratio.
    (d) Fitting a feature to the
    model in Eq.~\eqref{eq:intensity} yields
    estimates for $\vec{r}_p$, $a_p$ and $n_p$.}
    \label{fig:schematic}
\end{figure}

The custom-built 
holographic microscope used for Lorenz-Mie microscopy is
shown schematically in Fig.~\ref{fig:schematic}(a).
It illuminates the sample with a collimated laser 
beam whose electric field may
be modeled as a plane wave of frequency $\omega$ 
and vacuum wavelength $\lambda$
propagating along the $\hat{z}$ axis,
\begin{equation}
\label{eq:incidentfield}
\vec{E}_0(\vec{r}) = u_0 e^{i k z} e^{-i \omega t} \, \hat{x}.
\end{equation}
Here, $u_0$ is the field's amplitude and $k = 2\pi n_m/\lambda$
is the wavenumber of light in a medium
of refractive index $n_m$.
The beam is assumed to be linearly polarized along $\hat{x}$.
Our implementation uses a fiber-coupled diode laser
(Coherent Cube) operating at $\lambda = \SI{447}{\nm}$.
The \SI{10}{\milli\watt} beam is collimated at a diameter of
\SI{3}{\mm}, which more than fills the
input pupil of the microscope's objective lens
(Nikon Plan Apo, $100\times$, numerical aperture 1.4,
oil immersion).
In combination with a \SI{200}{\mm} tube lens, this
objective relays images to a grayscale camera
(FLIR Flea3 USB 3.0) with a \SI{1280 x 1024}{pixel}
sensor, yielding a system magnification of
\SI{48}{\nm\per pixel}.

A colloidal particle located at $\vec{r}_p$ 
relative to the center of the microscope's focal plane
scatters a small proportion
of the illumination
to position $\vec{r}$ in the focal plane of the microscope,
\begin{equation}
    \label{eq:scatteredfield}
    \vec{E}_s(\vec{r}) 
    = 
    E_0(\vec{r}_p) \, \vec{f}_s(k(\vec{r} - \vec{r}_p)).
\end{equation}
The scattered wave's relative amplitude, phase and polarization 
are described by the Lorenz-Mie scattering
function, $\vec{f}_s(k\vec{r})$, which generally depends on 
the particle's size, shape, orientation and composition \cite{bohren83,mishchenko02,gouesbet11}.
For simplicity, we model the particle as an isotropic
homogeneous sphere, so that $\vec{f}_s(k \vec{r})$
depends only on
the particle's radius, $a_p$, and its refractive
index, $n_p$.

The incident and scattered waves interfere in
the microscope's focal plane.
The resulting interference pattern is magnified by the
microscope and is relayed to the camera \cite{leahy2020large}, which
records its intensity.
Each snapshot in the camera's video stream
constitutes a hologram of the particles in the observation volume.
The image in Fig.~\ref{fig:schematic}(b) is
a typical experimentally recorded hologram of four colloidal silica
spheres.

The distinguishing feature of Lorenz-Mie microscopy is
the method used to extract information from 
recorded holograms.
Rather than attempting
to reconstruct the three-dimensional light
field that created the hologram,
Lorenz-Mie microscopy instead treats the
analysis as an inverse problem,
modeling the recorded intensity pattern as \cite{lee07a}
\begin{equation}
\label{eq:intensity}
I(\vec{r}) 
= 
u_0^2 \abs{\hat{x} + e^{i k z_p} \vec{f}_s(k (\vec{r} - \vec{r}_p))}^2 + I_0,
\end{equation}
where $I_0$ is the calibrated
dark count of the camera.
Fitting Eq.~\eqref{eq:intensity} to a measured hologram,
such as the examples in Fig.~\ref{fig:schematic}(c),
yields the ideal holograms shown in Fig.~\ref{fig:schematic}(d) together with values for
each particle's three-dimensional position, $\vec{r}_p$, 
as well as its
radius, $a_p$, and its refractive index, $n_p$, at the imaging wavelength.

Lorenz-Mie microscopy also
is implemented in a commercial holographic particle
characterization instrument (Spheryx xSight),
whose optical train differs significantly
from that of the custom-built instrument
and whose analytical software was developed
independently. We use a combination of
our own refined measurements and results
obtained with xSight to provide
experimental validatation for our
machine-learning implementation.

\subsubsection{Holographic optical trapping}
\label{sec:trapping}

Holographic optical traps 
with a vacuum wavelength of \SI{1064}{\nm} 
are projected into the sample
using the same objective lens that is used 
for holographic microscopy.
The traps are powered by a fiber laser 
(IPG Photonics YLR-10-LP) whose wavefronts
are imprinted with computer-generated
phase holograms \cite{grier03} using a liquid crystal
spatial light modulator (Holoeye Pluto).
The modified beam is relayed into the objective
lens with a dielectric multilayer
dichroic mirror (Semrock), which permits
simultaneous holographic trapping
and holographic imaging.

\subsection{Conventional Analysis}
\label{sec:conventional}

The first challenge in using Eq.~\eqref{eq:intensity}
to analyze a hologram is to detect
features of interest due to particles
in the field of view.
The concentric-ring pattern of a colloidal particle's hologram
can confound traditional object-detection algorithms
that seek out simply connected regions of similar intensity.
This problem has been addressed with
two-dimensional mappings such as circular Hough
transforms that coalesce
concentric rings into compact peaks
\cite{cheong09,parthasarathy12,krishnatreya14a}
that can be detected and localized with
standard peak-finding algorithms
\cite{crocker96}.
This approach is reasonably effective for detecting and localizing
holograms of well-separated particles.
It performs poorly for concentrated samples, however,
because overlapping scattering patterns create 
spurious peaks in the
transformed image that can trigger false positive detections.
These artifacts can be mitigated by
limiting the range over which rings
are coalesced at the cost of reducing
sensitivity to larger holographic features.
Optimizing the trade-off between 
false-positive and false-negative
detections requires tuning the search range in parameter space
and therefore creates a barrier to a fully-automated
implementation.

Having selected regions of interest such as the examples in
Fig.~\ref{fig:schematic}(c), the next step is to obtain
estimates for the particles' positions and properties
that are good enough to bootstrap nonlinear least-squares fits.
Reference implementations \cite{dimiduk16,lee07a} 
of Lorenz-Mie microscopy use the initial
localization estimate for the in-plane position
and wavefront-curvature estimates for $a_p$ and $z_p$. 
The initial value of $n_p$ often is based
on \emph{a priori} knowledge, which is undesirable for unattended
operation.

Ideally,
these initial stages of analysis should proceed
with minimal intervention, robustly identifying features and
yielding reasonable estimates for parameters over the widest
domain of possible values.
Applications that would benefit from real-time performance
also place a premium on fast algorithms, particularly those
that perform effectively on standard computer hardware.
These requirements can be satisfied with
machine-learning algorithms, which
surpass conventional algorithms in robustness, generality and
speed.

\subsection{Machine Learning Analysis}
\label{sec:machinelearning}

Previous efforts to streamline holographic particle characterization
with machine-learning techniques \cite{yevick14,hannel18,Schneider2016}, have addressed
separate facets of the problem, 
specifically feature localization \cite{hannel18,Newby2018}
and property estimation \cite{yevick14,Schneider2016}.
The former problem has been addressed with
convolutional neural networks (CNNs), 
which are widely used for
object localization \cite{hannel18,redmon2018YOLOv3}.
CNNs being far less common in regression applications,
the latter problem has been addressed 
by feeding features' radial intensity profiles 
into standard feed-forward neural networks \cite{Schneider2016}
or support vector machines \cite{yevick14}.
Although each effort has been successful in its domain,
combining them into an end-to-end analytical pipeline
has provided only modest
improvements in processing speed and robustness
because of the overhead involved in
extracting radial profiles and accounting for 
inevitable localization errors.

\begin{figure*}
    \centering
    \includegraphics[width=0.8\textwidth]{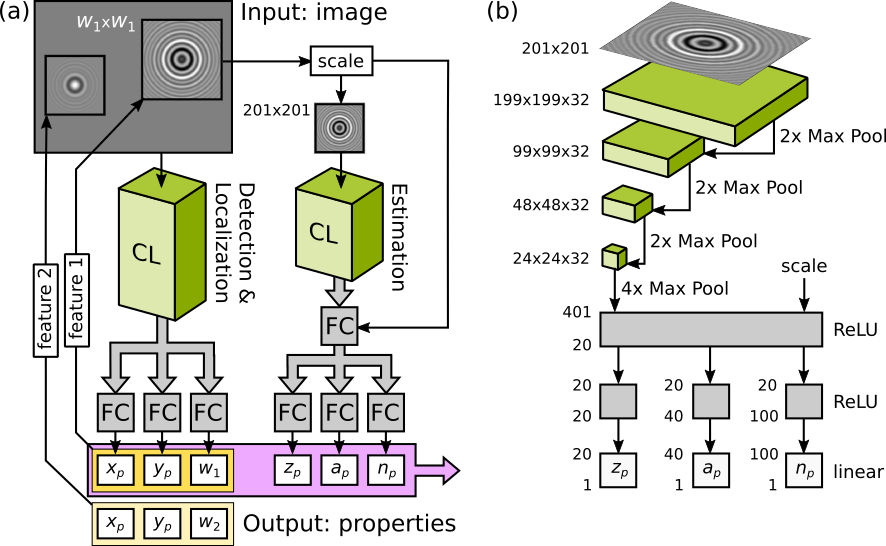}
    \caption{(a) CATCH uses a
    deep convolutional neural network (YOLOv3)
    to detect, localize and
    estimate the extent, $w_n$, of features in normalized holograms.
    Each feature is cropped from the image, scaled
    to a standard \SI{201x201}{pixel} format,
    and transferred to a second network
    that estimates the particle's axial position, radius and
    refractive index. Each network consists of convolutional
    layers (CL) that analyze image data and feed their results
    into fully connected (FC) layers that perform regression.
    (b) Detailed view of the estimation
    network. Four convolutional layers
    alternate with max-pooling layers
    to map the input image into a
    \num{400}-element vector that is
    concatenated with the image's scale factor.
    The resulting \num{401}-element vector
    is reduced to a \num{20}-element vector
    that describes the particle by
    a fully-connected layer with ReLU activation.
    The particle description is parsed into
    estimates for the axial position, $z_p$,
    particle radius, $a_p$, and refractive index,
    $n_p$, by three fully connected ReLU-activated
    layers feeding into three output layers
    with linear activation.
    }
    \label{fig:network}
\end{figure*}

We address the need for fast, fully automated
hologram analysis with a modular machine-learning
system based entirely on highly optimized deep convolutional
neural networks.
The system, shown in Fig.~\ref{fig:network},
is trained with synthetic data
that cover the entire anticipated domain of
operating conditions without requiring
manual annotation.
Each module yields useful intermediate
results, and the end-to-end system effectively
bootstraps full-resolution fits, which we
validate with experimental data.

The first module identifies features of interest
in whole-field holograms, localizes them,
and estimates their extents.
Each detected feature then is cropped from the image and
passed on to the second module, which estimates the
particle's radius, refractive index and axial position.
A feature's pixels and parameter estimates
then can be passed on to the third module, not
depicted in Fig.~\ref{fig:network}, which 
refines the parameter estimates by
performing a nonlinear least-squares fit to Eq.~\eqref{eq:intensity}.
This modular architecture permits limiting the analysis
to just what is required for the application at hand.

\subsubsection{Detection and Localization}
\label{sec:darknet}

The detection module is based on the darknet implementation
of YOLOv3, a state-of-the-art real-time object-detection framework
that uses a convolutional neural network to identify features of
interest in images, to localize them and, optionally, to classify them
\cite{redmon2018YOLOv3}.
Given our focus on detection and localization, we
adopt the comparatively simple and fast TinyYOLO variant,
which consists of \num{23} convolutional layers
with a total of \num{25620} adjustable parameters defining
convolutional masks and their weighting factors.

Taking a grayscale image as input, the model returns
estimates for each of the detected features'
in-plane positions, $(x_p, y_p)$, and their
extents.
These regions of interest can be used immediately
to measure particle concentrations, for example,
or they
can be passed on to
the next module for further analysis.

\subsubsection{Parameter Estimation}
\label{sec:keras}

CATCH estimates a particle's axial position, $z_p$, radius,
$a_p$, and refractive index, $n_p$,
by passing the associated block of pixels through
a second deep convolutional neural network
for regression analysis.
The regression network, depicted schematically in
Fig.~\ref{fig:network}(b),
consists of \num{19} layers 
with a total of \num{34983} trainable parameters
and is constructed with the open-source Tensorflow framework 
\cite{tensorflow2015-whitepaper}
using the Keras application programming interface (API).
The network's input is a block of pixels cropped from
the holographic image and then scaled down 
by an integer factor to
\SI{201 x 201}{pixel}.
Scaling enables the estimator
to accommodate scattering patterns with a wide range of extents in the camera plane.

The scaled image data initially pass 
through a series of convolutional
and pooling layers that reduce the dimensionality of the
regression space. 
The flattened output then is fed through a shared 
fully-connected layer along with the image's scale
factor to produce a \num{20}-element vector
that describes the particle's position in parameter space.
This layer uses rectified linear activation units
(ReLU) whose nonlinear response enables the network to learn 
complicated functions and whose near-linear form
facilitates rapid training
\cite{pmlr-v15-glorot11a}.
The output of this layer is decoded by three independent
ReLU-activated layers whose responses 
are scaled into dimensional values for $z_p$, $a_p$ and $n_p$ by linear output layers.

\subsubsection{Training}
\label{sec:training}

Convolutional neural networks have the capacity
to uncover low-dimensional approximate solutions
to information-processing problems characterized
by large numbers of internal degrees of freedom \cite{Rotskoff2018}.
To learn these patterns, however, the
network must be trained
with data that span
the parameter range of interest at the
desired resolution.
Defining $R(p_j)$ to be the range of the
output parameter
$p_j$ in a set of $M$ coupled parameters
and $\Delta p_j$ to be the desired
resolution in that parameter, naive scaling
suggests that the number of elements required
for a comprehensive
training set should satisfy
\begin{equation}
    N \leq \prod_{j=1}^M\frac{R(p_j)}{\Delta p_j}.
    \label{eq:rule}
\end{equation}
The upper limit corresponds to sampling every
possible solution,
assuming that the generative function does not
vary substantially over the range $\Delta p_j$.
Smaller training sets suffice for problems that
have an inherently lower-dimensional
underlying structure.
Because
the two components of the CATCH network
address different aspects of
hologram analysis, we train them separately,
thereby reducing the dimensionality of the
overall problem and helping to ensure
rapid and effective convergence with a reasonable
amount of training.

We train the detection and localization network to
recognize features in holographic microscopy images
using a custom data set consisting of $N = \num{10000}$ 
synthetic holographic images
for training and an additional \num{1000} images for 
validation.
If we assume that $x_p$ and $y_p$ are the only
relevant output parameters, then this number of
images should provide no worse than 
$\Delta x_p = \Delta y_p \approx \SI{10}{pixel}$
localization resolution for
\SI{1280x1024}{pixel} images
according to Eq.~\eqref{eq:rule}.

The synthetic images are designed to mimic experimental 
holograms
over the anticipated domain of operation, with between
zero and five particles positioned randomly in the field 
of view.
Particles are assigned radii between 
$a_p = \SI{200}{\nm}$ 
and $a_p = \SI{5}{\um}$
and refractive indexes between $n_p = \num{1.338}$ and 
$n_p = \num{2.5}$,
and are located along the optical axis at distances from 
the focal plane between 
$z_p = \SI{50}{pixels}$ and $z_p = \SI{600}{pixels}$,
with each axial pixel corresponding to the 
in-plane scale of \SI{48}{\nm}.
The extent, $w_p$, 
of each holographic feature is defined 
to be the diameter enclosing \num{20} interference 
fringes and therefore
scales with the particle's size and axial position.
Ideal holograms computed with Eq.~\eqref{eq:intensity}
are degraded with \SI{5}{\percent} additive
Gaussian noise.
The ground truth for localization training consists of 
the in-plane coordinates, $(x_p, y_p)$, 
of each feature in a hologram
together with the features' extents, $w_p$.
We trained for \SI{500000}{epochs} with a batch size of 
\num{64} images and 
\num{32} subdivisions. 

We train the estimator network on a second set of
$N = \num{10000}$ synthetic single-particle
holograms with a validation set of \num{1000} images,
covering the same parameter range used to train
the localization network.
In addition to adding
\SI{5}{\percent} Gaussian noise to the
intensity pattern, we also
incorporate up to 
\SI{10}{pixel} localization error 
and \SI{10}{\percent} error in extent
to simulate worst-case performance by the
first stage of analysis.
The network is trained with the Adam optimizer
\cite{kingma2014adam}
for \num{5000} epochs with a batch size of
\num{64} images
using minimal dropout and L2 regularization to 
prevent overfitting.
Naive application of Eq.~\eqref{eq:rule}
then suggests that we should expect
$\Delta z_p \leq \SI{1.2}{\um}$, 
$\Delta a_p \leq \SI{0.22}{\um}$ and
$\Delta n_p \leq \num{0.05}$.


\section{Results and Discussion}
\label{sec:results}

\subsection{Validation with Synthetic Data}
\label{sec:synthetic}

The CATCH network's performance is validated
first with synthetic data
and then through experiments on model systems.
The synthetic validation data set
consists of \num{10000} holograms
that were generated independently of the
data used for training.
This data set is designed to assess
performance under ideal imaging conditions without
the additional complexity of overlapping features 
in multi-particle holograms.
Each synthetic hologram contains one particle with randomly
selected properties placed at random
in the \SI{1280x1024}{pixel} field of view 
and includes \SI{5}{\percent} additive Gaussian noise.

\subsubsection{Processing Speed}
\label{sec:speed}

Tests were performed with hardware-accelerated
versions of each algorithm running on a desktop
workstation equipped with an NVIDIA Titan Xp GPU.
On this hardware, conventional algorithms 
\cite{cheong09,krishnatreya14a,crocker96,allan16trackpy,dimiduk16}
perform a complete
end-to-end single-frame analysis in roughly \SI{1}{\second}.
By contrast, 
the CATCH network's
detector and localizer requires \SI{20}{\ms}
per frame and
the machine-learning estimator requires an additional
\SI{0.9}{\ms} per feature.
This is fast enough for real-time performance
assuming a typical frame rate of \SI{30}{frames\per\second}.

\subsubsection{Detection Accuracy}
\label{sec:detection}

\begin{figure*}
    \centering
    \includegraphics[width=\textwidth]{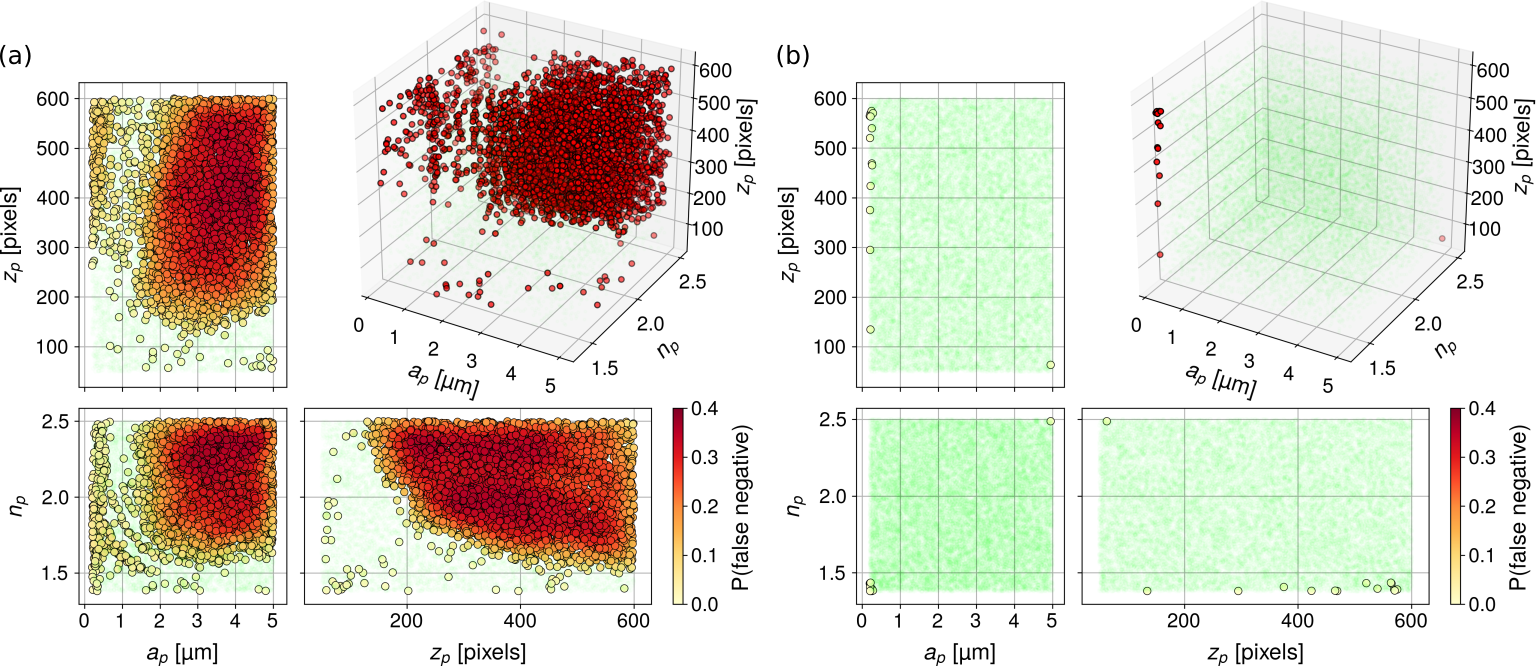}
    \caption{False negative detections in simulated holograms plotted
      as filled circles and colored by
    the local error probability.
    (Green) points denote correct 
    positive detections. (a) Conventional
    feature detection algorithms
    miss up to \SI{40}{\percent} of particles in 
    \num{10000} simulated
    single-particle holograms. (b) The convolutional neural network
    misses fewer than
    \SI{0.1}{\percent} of the \SI{25000} plotted 
    features over
    the same parameter range.}
    \label{fig:falsenegative}
\end{figure*}

When assessing detection accuracy, 
we are concerned primarily with the rate of false negative
detections. 
False positive detections are less concerning because
they can be identified and filtered through post-processing, but
false negatives represent lost information.
Conventional feature detection algorithms
\cite{cheong09,parthasarathy12,krishnatreya14a}
have been shown to work well for small, weakly-scattering particles.
Over the larger range of parameter space plotted
in Fig.~\ref{fig:falsenegative}(a), however, conventional algorithms
fail to detect up to \SI{40}{\percent} of particles, even under ideal conditions.
Over the same range, the neural network misses fewer than \SI{0.1}{\percent}
of features, as shown in Fig.~\ref{fig:falsenegative}(b),
and proposes no false positives.
The false negatives occur for very small particles that are
nearly index matched to the medium whose holograms have
the poorest signal-to-noise ratios in this study.

This dramatic improvement in detection reliability
greatly expands the parameter space for unattended
Lorenz-Mie particle characterization.
It allows for automated analysis of larger volumes,
larger particles and larger ranges of particle characteristics
in a single sample.
Such systems could have been analyzed previously, but
would have required human intervention.

\subsubsection{Localization and Feature Extent}
\label{sec:localization}

\begin{figure}
    \centering
    \includegraphics[width=0.9\columnwidth]{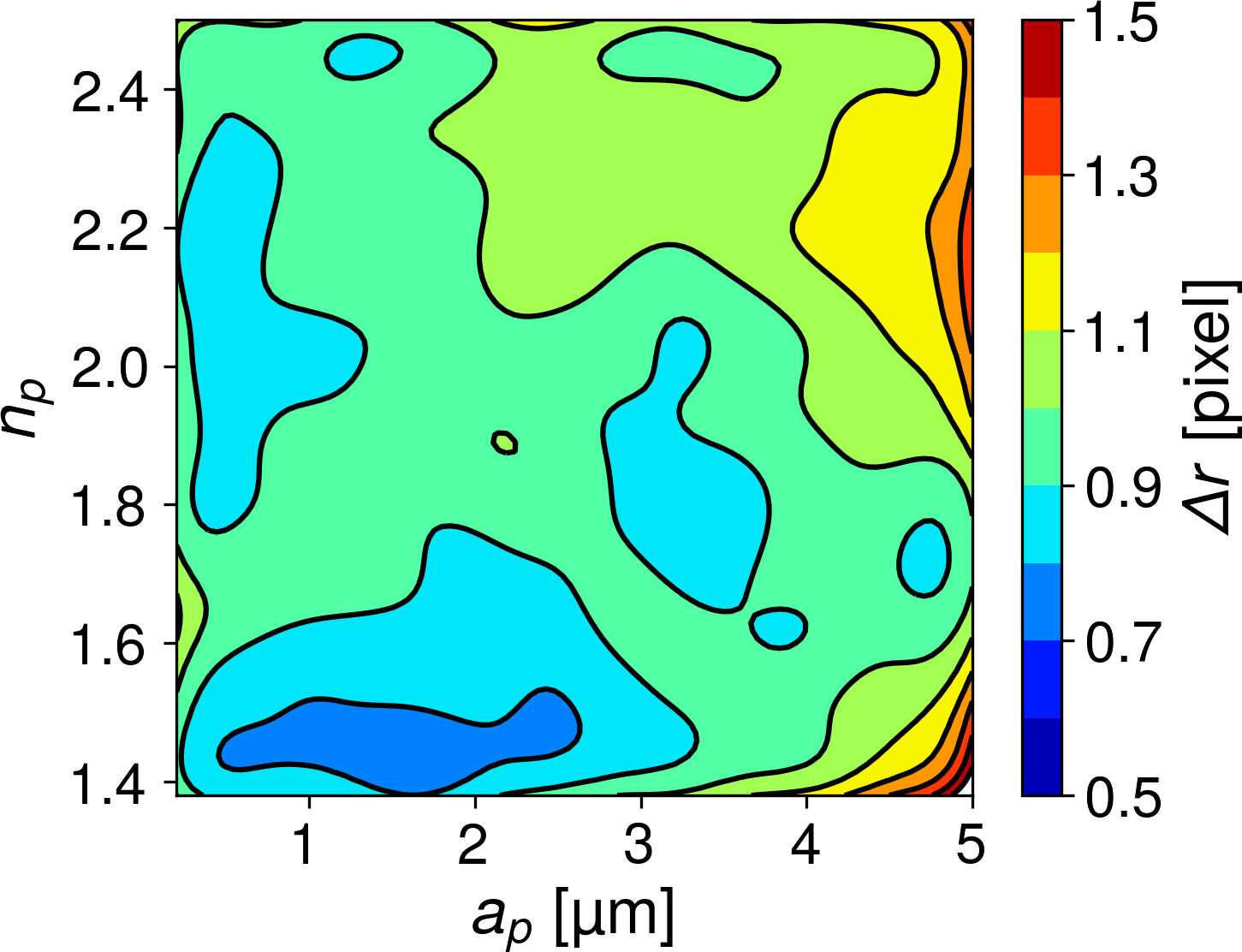}
    \caption{In-plane localization error, $\Delta r$, 
    as a function 
    of particle radius and refractive index, averaged over
    axial position, for all \num{24994} true positive detections.
    Obtained from TinyYOLO implementation of the network localizer.}
    \label{fig:localization}
\end{figure}

Localization accuracy is assessed for true-positive detections
on synthetic images
using the input particle locations as the ground truth.
As presented in Fig.~\ref{fig:localization},
the net in-plane localization error is
smaller than 
$\Delta x_p = \Delta y_p = \SI{1.5}{pixel}$,
or \SI{70}{\nm},
across the entire range of parameters, and typically is
better than \SI{1}{pixel}.
The localizer therefore outperforms the 
naive estimate for localization precision
in Eq.~\eqref{eq:rule}, presumably because
the CNN has identified a low-dimensional
representation for the problem.
Estimates for the features' extents, $w_p$, vary
from the ground truth by \SI{15}{\percent}
with a bias toward underprediction.
This figure of merit is a target for future
improvement because scaling errors
propagate into the regression analysis and
are found to increase errors in the estimates for
$a_p$ and $z_p$.

\subsubsection{Characterization}
\label{sec:characterization}

\begin{figure*}
    \centering
    \includegraphics[width=\textwidth]{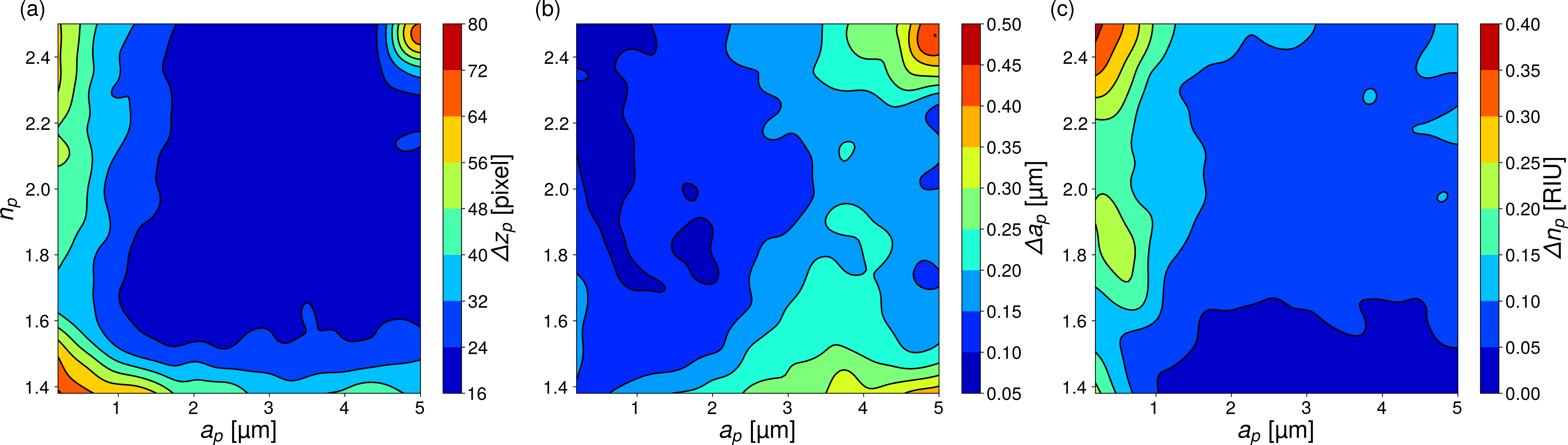}
    \caption{Root-mean-square errors in
    (a) axial position, $\Delta z_p$, (b) radius, $\Delta a_p$ 
    and (c) refractive index, $\Delta n_p$, as a function
    of radius and refractive index on a set of \num{25000} cropped holograms. 
    Results are averaged over placement errors.}
    \label{fig:estimation}
\end{figure*}

Figure~\ref{fig:estimation} summarizes the
regression network's performance for estimating 
axial position, particle size and refractive index
in the validation set of synthetic data.
Each panel shows the root-mean-square error for one
parameter as a function of $a_p$ and $n_p$,
averaged over $\vec{r}_p$. Over most of the
parameter domain, the estimator predicts
the relevant parameter to within \SI{10}{\percent}.
This is not quite as good as the naive estimate
from Eq.~\eqref{eq:rule} proposes, which
suggests that the parameter space has not
been sampled finely enough to resolve 
the structure of the underlying Lorenz-Mie
scattering problem. Achieving the
part-per-thousand precision offered
by nonlinear least-squares fits would require
on the order of $N = \num{e9}$ 
images, according to naive scaling.

Conventional gradient-descent
fits to the Lorenz-Mie theory
display pronounced anticorrelations between $a_p$ and $n_p$
\cite{ruffner2018lifting}.
No strong cross-parameter correlation is evident in the error
surfaces plotted in Fig.~\ref{fig:estimation}.
This difference highlights a potential benefit
of machine-learning regression for complex image-analysis
tasks.
Unlike conventional fitters, 
machine-learning algorithms do not attempt to follow
smooth paths through complex error landscapes, but rather
rely upon an internal representation of the error
landscape that is built up during training.
Directly reading out an optimal solution from such
an internal representation is computationally efficient
and less prone to trapping in local minima of the 
conventional error surface. 
Most importantly, unsupervised parameter estimation
eliminates the need for \emph{a priori} information
or human intervention in colloidal materials characterization.

\subsection{Validation with Experimental Data}
\label{sec:experimental}

Having validated the CATCH system's performance with synthetic 
data, we use it to analyze experimental data.
Some applications, such as measuring 
particle concentations, can be undertaken with
the detection and localization module alone.
Some other tasks require characterization data
and can be performed
with the output of the estimation module.
Still others use
machine-learning estimates to bootstrap nonlinear least-squares fits
to Eq.~\eqref{eq:intensity}.
The full end-to-end mode of operation benefits from the speed 
and robustness of machine-learning estimation and
delivers the precision of nonlinear optimization
\cite{lee07a,krishnatreya14}.

\subsubsection{Fast and accurate colloidal concentration measurements}
\label{sec:concentration}

CATCH's
detection subsystem rapidly counts particles
passing through the microscope's observation volume
and thus can measure their concentration.
Its ability to detect particles over a large axial range
is an advantage relative to conventional image-based
particle-counting techniques 
\cite{ripple2016correcting}, 
which have a limited
depth of focus and thus a more restricted observation
volume.

Although the holographic microscope's
measurement volume might be known
\emph{a priori}, CATCH also can estimate
the  effective observation volume
from the least bounding rectangular
prism that encloses all detected particle locations.
This internal calibration is
most effective for particles
that remain dispersed throughout the height of the
channel.
For such samples, this protocol addresses uncertainties
due to variations in actual
channel dimensions and accounts for detection limits
near the boundaries of the observation volume.

We demonstrate machine-learning
concentration measurements
on a heterogeneous
sample created by mixing four different
populations of monodisperse colloidal spheres:
two sizes of polystyrene spheres
(Thermo Scientific, catalog no.\ 5153A, 
$a_p = \SI{0.79}{\um}$; 
Duke Standards, catalog no.\ 4025A,
$a_p =  \SI{1.25}{\um}$), and
two sizes of silica spheres
(Duke Standards, catalog no.\ 8150,
$a_p = \SI{0.79}{\um}$;
Bangs Laboratories, catalog no.\ SS05N,
$a_p = \SI{1.15}{\um}$).
Each population of spheres is dispersed in water
at a nominal concentration of
\SI{4e6}{particles\per\milli\liter}.
Equal volumes of these monodisperse stock dispersions
are mixed to create the heterogeneous sample.
Such mixtures can pose challenges for conventional
techniques such as dynamic light scattering, which 
assume that the scatterers are drawn from a unimodal
distribution.
No other particle-characterization technique would
be able to differentiate particles with similar
sizes but different compositions.

A \SI{30}{\micro\liter} aliquot of the four-component
dispersion is introduced into a channel formed by bonding
the edges of a \#1.5 cover glass to the face of a
glass microscope slide with UV-cured adhesive
(Norland Products, catalog no.\ NOA81).
The resulting channel is roughly \SI{1}{\mm} wide, \SI{2}{\cm} long
and \SI{15}{\um} deep.
Once the cell is mounted on the stage of the holographic
microscope,
we transport roughly \SI{10}{\micro\liter} of this
sample through the microscope's observation volume
at roughly
\SI{1}{\milli\meter\per\second}
in a capillary-driven flow.
A data set of 
\num{47539} video frames recorded
over \SI{26}{\minute}
probes a total volume of
\SI{0.83(14)}{\micro\liter}
given the effective observation volume of
\SI{25x38x18(3)}{\um},
or \SI{17(3)}{\pico\liter}.
The \SI{18}{\percent} uncertainty in the axial extent
dominates the uncertainty in the effective observation volume. 
In-plane dimensions are determined to better than
\SI{1}{\percent}.

The CATCH detection module
reports \num{2967} features in this sample,
which corresponds
to a net concentration of
\SI{3.6(6)e6}{particles\per\milli\liter}.
This value is consistent with
expectations based on the concentrations
of the stock dispersions and agrees reasonably well
with the value of
of \SI{3.1e6}{particles\per\milli\liter}
obtained with xSight.
CATCH is fast enough to complete
the concentration estimate in the
time required to record the images.

\subsubsection{Characterizing heterogeneous dispersions}
\label{sec:mixture}

\begin{figure*}
    \centering
    \includegraphics[width=0.85\textwidth]{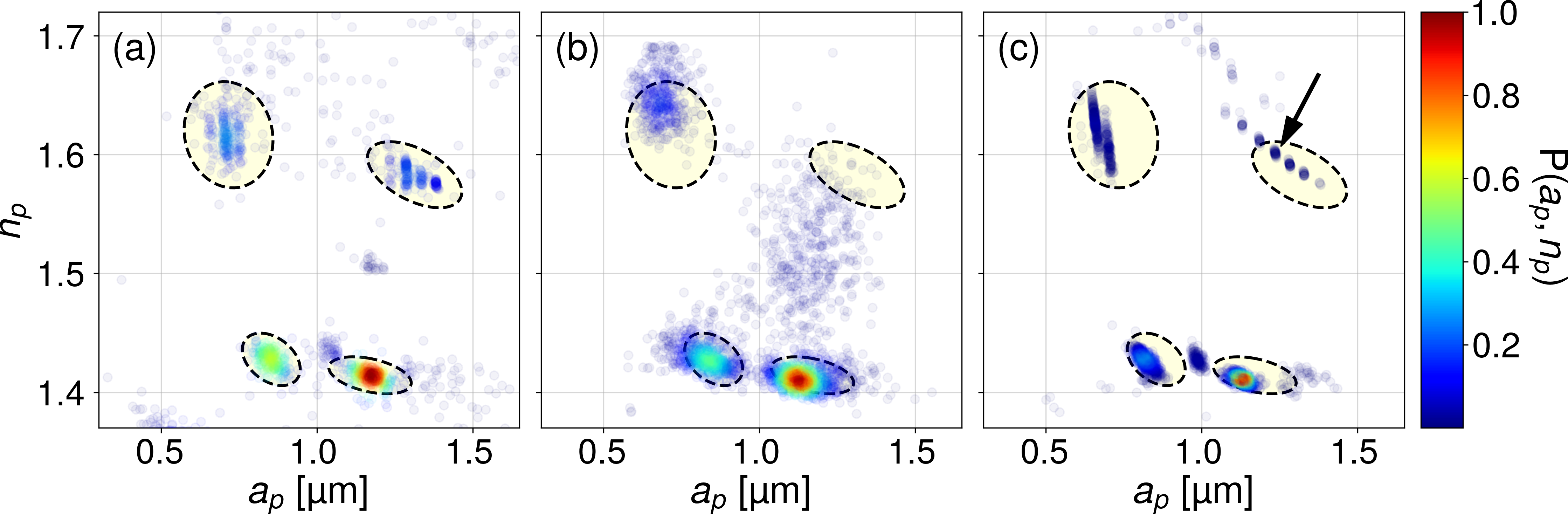}
    \caption{Measurements of the radius, $a_p$, 
    and refractive index, $n_p$, of a mixture of four monodisperse populations of polystyrene 
    and silica spheres. Each point represents the
    properties of a single particle and is colored by the relative probability density of observations, $P(a_p, n_p)$.
    (a) Properties of \num{1917} particles reported
    by xSight.
        Ellipses represent \SI{99}{\percent} confidence intervals for each
    population of particles.
    (b) Predictions for another \num{2967} particles
    recorded on the custom-built microscope and
    analyzed by the CATCH convolutional neural network.
    (c) CATCH predictions refined by nonlinear
    least-squares fits to Eq.~\eqref{eq:intensity}.
    The arrow indicates the globally 
    optimal characterization
    parameters for the large polystyrene spheres in
    this system.}
    \label{fig:mixture}
\end{figure*}

The detection subsystem is not trained to
distinguish among different
types of particles.
The estimation subsystem, however,
can differentiate particles both by size and 
also by refractive index.
The scatter plots 
in Fig.~\ref{fig:mixture} show
holographic particle characterization data
of thousands of particles from the four-component dispersion described in the previous section.
Points are colored by the 
relative density of observations,
$P(a_p, n_p)$.

The results plotted in Fig.~\ref{fig:mixture}(a)
are obtained with xSight and will be treated as
the ground truth. 
Dashed ellipses superimposed on the particle-resolved 
characterization data represent \SI{99}{\percent}
confidence intervals obtained
with principal component analysis for each of the
four populations of particles.
Additional data points outside these regions correspond
to impurity particles such as dimers 
as well as a small number of spurious results
caused by overlapping holograms.
The same ellipses are superimposed on the
results obtained with CATCH in Fig.~\ref{fig:mixture}(b)
and on the refined estimates bootstrapped by CATCH
in Fig.~\ref{fig:mixture}(c).

The upper two ellipses centered on
refractive index around \num{1.60}
correspond to the two
sizes of polystyrene
spheres in the mixture.
The two lower ellipses correspond to the silica spheres
with a refractive index around \num{1.40}.
xSight clearly distinguishes the 
two compositions of spheres 
by their refractive
indexes.
The ability to differentiate particle populations
both by size and by composition is a unique advantage
of holographic particle characterization
relative to all other particle characterization
technologies.

Figure~\ref{fig:mixture}(b) shows results from a separate
measurement on the same colloidal sample performed with the
custom-built holographic video microscope and analyzed
with CATCH.
All four populations are visible in the scatter plot,
and the silica characterization results agree
quantitatively with xSight measurements.
The larger polystyrene spheres appear as a 
poorly localized cloud of points because 
that range of parameter space is characterized
by a large number of nearly degenerate solutions
\cite{ruffner2018lifting}.

Although CATCH does not achieve the full precision
of the Lorenz-Mie analysis, its results still are
close enough to the ground truth to bootstrap
nonlinear least-squares fits.
The results in Fig.~\ref{fig:mixture}(c) show the
same data from Fig.~\ref{fig:mixture}(b) after
nonlinear fitting. 
The predictions for all four
populations are consistent 
with xSight measurements, albeit
with systematic offsets that likely
arise from differences in the two instruments'
optical trains that are not accounted for by
the model in Eq.~\eqref{eq:intensity}
\cite{leahy2020large}.
The root-mean-squared displacements of
the CATCH estimates from the corresponding
refined values,
$\Delta a_p = \SI{89}{\nm}$ and 
$\Delta n_p = \num{0.04}$,
are consistent with the errors estimated
with synthetic validation data in Sec.~\ref{sec:synthetic}.

Results for the larger polystyrene
spheres do not converge into a well-defined 
cluster, but rather form a series of islands
that constitute a set of nearly degenerate
solutions \cite{ruffner2018lifting}.
The same structure also can be discerned in
xSight results. The central island,
indicated by an arrow in Fig.~\ref{fig:mixture}(c),
appears to correspond 
to the globally optimal solution based on 
the chi-squared statistic for all fits.
Neither the Levenberg-Marquardt least-squares
optimizer nor the Nelder-Mead simplectic search
algorithm consistently converges to this 
particular solution starting from the 
CATCH estimates for these particles.
Even for this challenging case, however, 
the parameters proposed by CATCH converge
to reasonable values within
the expected confidence interval,
which demonstrates that they are good enough
for practical applications.

\subsubsection{Tracking confined sedimentation}
\label{sec:tracking}

\begin{figure}
    \centering
    \includegraphics[width=0.9\columnwidth]{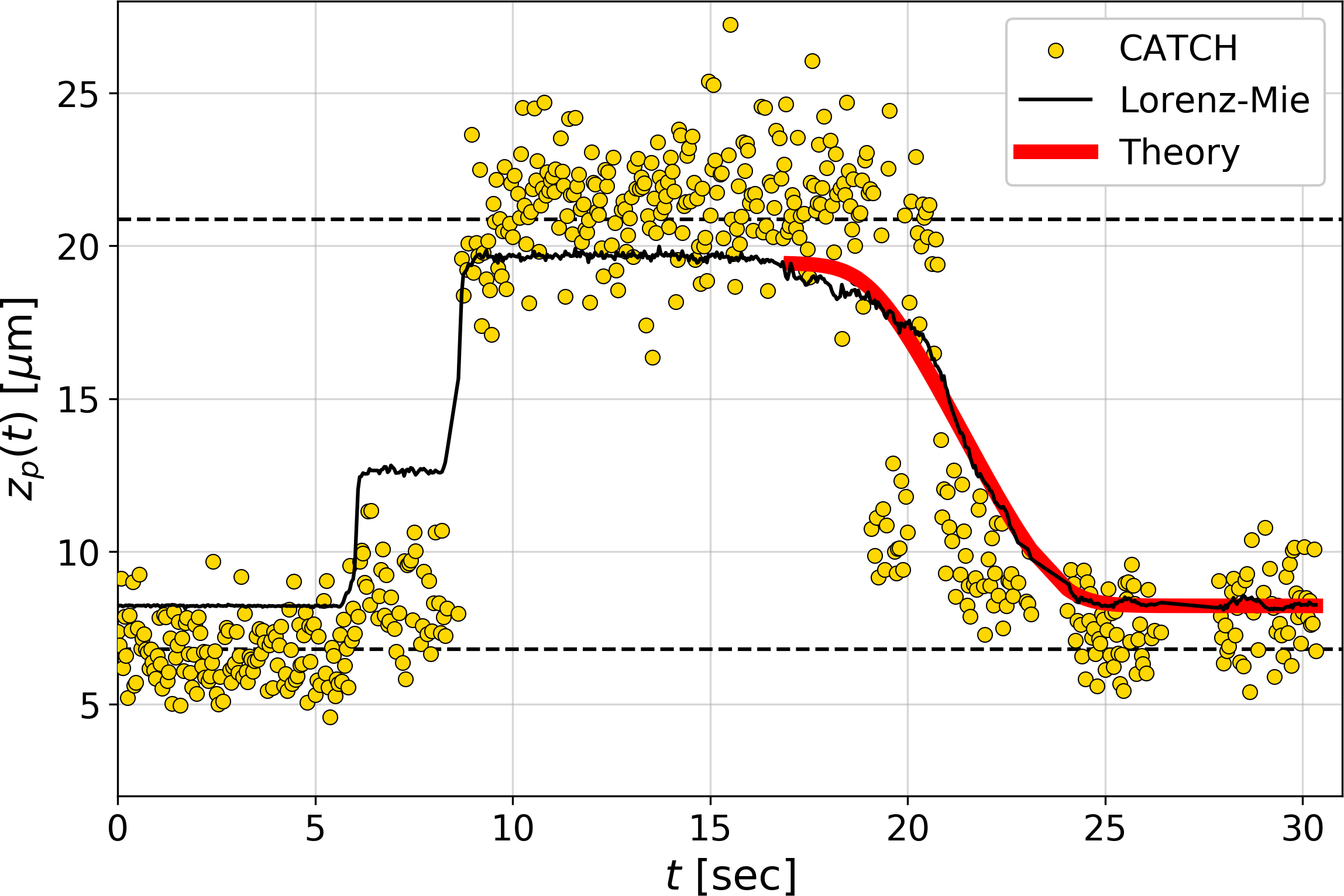}
    \caption{Estimated (points) and refined 
    (solid black curve) axial trajectory of a colloidal silica sphere
    being lifted to the upper wall of a
    water-filled channel and allowed to 
    sediment to the lower wall under gravity. The heavy
    (red) curve is a fit to Eq.~\eqref{eq:sedimentationmodel}
    for the density of the particle
    and the positions of the walls,
    which
    are indicated by horizontal dashed lines.}
    \label{fig:sedimentation}
  \end{figure}
  
To illustrate three-dimensional particle
tracking based on CATCH estimation,
we measure the sedimentation of
a colloidal sphere between two
parallel horizontal surfaces.
The influence of slot confinement
on a colloidal sphere's
in-plane drag coefficient has been
reported previously using
conventional imaging
\cite{dufresne01}.
The axial drag coefficient has
not been reported, presumably because
of the difficulty of measuring the
axial position with sufficient accuracy.

We perform the measurement on a 
colloidal silica sphere (Bangs Laboratories, catalog no.\ SS05N)
dispersed in \SI{30}{\micro\liter}
of deionized water that is contained in a glass sample chamber 
formed by bonding the edges of a glass cover slip to the face 
of a glass microscope slide.
Holographic optical traps are projected into the sample
using the same objective lens that is used to record holograms
\cite{grier03}.
We lift the sphere
to the top of its sample chamber
using a holographic optical trap
\cite{grier03,obrien19} 
and then release it.
Analyzing the particle's trajectory 
then yields an estimate for 
the buoyant mass density
that can be compared with
an orthogonal estimate based on
the particle's holographically 
measured size and refractive index.

The discrete data points in Fig.~\ref{fig:sedimentation} are machine-learning
estimates of the particle's axial position, $z_p(t)$, as a function
of time, recorded at \SI{24}{frames\per\second}.
The solid (black) curve is obtained by fitting the sphere's
hologram to Eq.~\eqref{eq:intensity} starting from
machine-learning estimates for $\vec{r}_p(t)$, $a_p$ and $n_p$.
These fits converge to 
$a_p = \SI{1.14(4)}{\um}$ and
$n_p = \num{1.398(5)}$, which
are consistent with the manufacturer's 
specification and with the population-averaged
properties, $a_p = \SI{1.17(15)}{\um}$ and 
$n_p = \num{1.42(2)}$,
obtained with xSight.
The root-mean-square axial tracking error for this
data set is $\Delta z_p = \SI{2.8}{\um}$, 
which is consistent with errors estimated in Sec.~\ref{sec:synthetic}.

In addition to being acted upon by gravity, the particle also is
hydrodynamically coupled to the walls of its sample chamber,
which reduces its mobility.
The particle sediments under gravity at a rate,
\begin{subequations}
\label{eq:sedimentationmodel}
\begin{equation}
    \label{eq:sedimentation}
    \frac{d z_p}{dt} = 
    - \frac{4}{3} \pi a_p^3 
    \, (\rho_p - \rho_m) \, g \, \mu(z_p),
\end{equation}
that depends on the difference between
its mass density, $\rho_p$, and the
mass density of the medium, $\rho_m$.
Hydrodynamic coupling to the parallel glass walls
reduces the sphere's mobility, $\mu(z_p)$, by an amount
that depends on its axial position
within the channel.
Specifically, the flow field due to
the sedimenting sphere is modified by
no-slip boundary conditions at the lower and upper walls,
which are located
at $z = z_0$ and $z = z_0 + H$
relative to the microscope's focal plane, respectively.
For simplicity, we model the resulting
dependence by combining
lowest-order single-wall corrections
\cite{happel91}
with the Oseen linear superposition approximation
to obtain
\begin{equation}
    \label{eq:mobility}
    \mu(z) 
    \approx
    \frac{1}{6 \pi \eta a_p}
    \left( 1 
    - \frac{9}{8} \frac{a_p}{z - z_0} 
    - \frac{9}{8} \frac{a_p}{H - z + z_0}\right),
\end{equation}
\end{subequations}
where $\eta = \SI{0.89}{\milli\pascal\second}$ is the viscosity of water.
The solid (red) curve in Fig.~\ref{fig:sedimentation} is a fit of the
refined data (black curve)
to the prediction of Eq.~\eqref{eq:sedimentationmodel} with
$z_0$, $H$ and $\rho_p$ as adjustable parameters.
This fit yields 
$\rho_p = \SI[parse-numbers=false]{2.18\substack{+0.07\\-0.20}}{\gram\per\cubic\cm}$
assuming $\rho_m = \SI{0.997}{\gram\per\cubic\cm}$ for
water.

The particle's comparatively low mass density is consistent
with its low refractive index.
Maxwell Garnett effective medium theory \cite{markel16}
suggests that the particle's density
may be estimated from its refractive
index as
\begin{equation}
    \label{eq:maxwellgarnett}
    \rho_p = 
    \rho_0
    \frac{L_m(n_p) - L_m(1)}{L_m(n_0) - L_m(1)},
\end{equation}
where $\rho_0 = \SI{2.20}{\gram\per\cubic\cm}$ is the density of fused silica,
$n_0 = \num{1.465}$ is the refractive
index of fused silica at the imaging
wavelength, and
\begin{equation}
    \label{eq:lorentzlorenz}
    L_m(n) 
    = 
    \frac{n^2 - n_m^2}{n^2 + 2 n_m^2}
\end{equation}
is the Lorentz-Lorenz function.
The result,
$\rho_p = \SI{1.90(10)}{\gram\per\cubic\cm}$,
is consistent with the lower bound
of the kinematic estimate,
and so helps to validate \cite{krishnatreya14}
the accuracy and precision with
which CATCH characterizes and tracks
colloidal particles.

\section{Conclusions}
\label{sec:conclusions}

CATCH is an
end-to-end machine-learning
system for analyzing the properties of
colloidal dispersions from holographic microscopy images.
Based on YOLO and a custom-designed deep convolutional
neural network, this system delivers the full characterization
and tracking power of Lorenz-Mie microscopy with greatly improved
speed and robustness.
This implementation has been validated both with simulated
data and also through experimental measurements on model
colloidal dispersions.
These measurements illustrate the utility of CATCH for
measuring the concentrations of colloidal dispersions,
for characterizing the particles in heterogeneous
dispersions, and for measuring single-particle dynamics.

More generally, CATCH embodies a paradigm shift in
measurement theory, with machine-learning algorithms
replacing physical mechanisms and physics-based models
in precision measurements. The availability of such
``brain-in-a-box" instruments increases the speed
and robustness of such measurements and also promises
access to physical phenomena that cannot readily be measured
by other means.

Our open-source implementation 
of the end-to-end CATCH system is available online 
at \url{https://github.com/laltman2/CNNLorenzMie}.

\section*{acknowledgement}
This work was supported primarily by the MRSEC program of the
National Science Foundation under Award Number DMR-1420073.
Additional support was provided by the SBIR program of
the National Institutes of Health under Award Number
R44TR001590.
The Titan Xp GPU used for this work was provided by a GPU Grant from NVIDIA.
The Spheryx xSight holographic characterization 
instrument was acquired by
the NYU MRSEC as shared instrumentation.
The custom-built holographic trapping and microscopy
system
was developed with support from the MRI program of the NSF
under Award Number DMR-0922680.

%

\end{document}